\documentclass{xray_symp}
\usepackage{graphics,psfig}
\def\sbu{${\rm mag\,\,arcsec^{-2 }}$\ }

\def\ergsec{${\rm erg\,\,s^{-1}}$}
\newcommand{\mima}[2]{$_{\scriptscriptstyle -#1}^{\scriptscriptstyle +#2}$}

\def\hi{H\,{\small I}}
\def\ha{H$\alpha$}
\def\hn2{H$\alpha$+{[}N{\small II}{]}\  }

\def\lx{$\log(L_{\rm X})$}

\def\lxb{$\log(L_{\rm X}/L_{\rm B}$)}

\def\lirb{$\log(L_{\rm FIR}/L_{\rm B}$)}
\def\vv{\vspace*{-0.4mm}}
\begin{document}
\title{X-rays from Blue Compact Dwarf Galaxies}
\author{P.\ Papaderos \and K.J.\ Fricke}  
\institute{Universit\"{a}ts-Sternwarte G\"{o}ttingen, 
Geismarlandstra\ss e 11, 37083 G\"{o}ttingen, Germany}
\maketitle

\begin{abstract}
We present first results of a study devoted to the analysis of the
X-ray emission as tracer of activity and diagnostic tool for 
the structural properties of Blue Compact Dwarf Galaxies. 
The case of Henize\ 2-10 is being reviewed, 
and first trends are discussed.
\end{abstract}
% ----------------------------------------------------------------------------
\section{Introduction}
Blue Compact Dwarf Galaxies (BCDs) are metal poor ($Z\!\sim\! 1/10\,Z_{\odot}$)
and gas-rich extragalactic systems undergoing brief ($<10^7$ yr) bursts of star 
formation, separated by long ($\sim\!1$ Gyr) quiescent phases 
(see Thuan 1991 for a review). 
The detection of an extended and old low-surface-brightness (LSB) 
stellar component underlying the regions of active star formation 
in the majority of BCDs (Loose\ \&\ Thuan 1986, Kunth et al.\ 1988) 
implies that such systems are old gas-rich dwarf galaxies 
undergoing recurrent activity phases.
This faint, however dynamically important, mass 
constituent in BCDs, together with the \hi-gas and 
Dark Matter, determines the gravitational
potential within which starbursts occur. 
The physical origin of starbursts in the mostly isolated BCDs 
is not yet understood.
Hypotheses put forward invoke a dynamical perturbation by a nearby
\hi-cloud (Taylor et al. 1994), interaction/merging with another dwarf 
galaxy (Comte et al. 1996), or inflow and collapse of their massive 
\hi-halo onto the stellar LSB-component (Loose \& Fricke 1981). 
The prime intrinsic properties of a BCD controlling the morphology, 
the spatial extent, and the strength of the burst are still a
matter of debate. 
Among the hypotheses proposed are a massive, BCD-specific, Dark Matter 
halo dominating entirely the mass (Meurer et al. 1998).  
In another scenario the mass and the structural properties 
of the stellar LSB-component, i.e. the shape of the gravitational 
potential of the underlying old stellar population, are 
believed to regulate the global star-formation process 
(Papaderos et al. 1996).

A further issue is related to the evolutionary links between BCDs and 
other classes of galaxies such as dwarf irregulars dIs and dwarf ellipticals dEs.
Papaderos et al. (1996) found that, at equal B-luminosity, the structural
properties of the underlying host galaxy of BCDs do systematically differ 
from dIs/dEs: at M$_{\rm B}$=--16 mag the central surface brightness and 
the exponential scale length of the LSB-component of BCDs are   
by $\sim 1.5$ mag brighter and by factors $\sim 2$ smaller than 
dIs/dEs. 
This result, corroborated in the range of more luminous BCDs by 
Marlowe et al. (1997), implies that the commonly accepted
evolutionary connection between BCDs and dIs cannot be maintained 
unless the LSB component of BCDs can undergo dynamical changes such 
as expansion and contraction on time scales of few $10^8$ yr 
(Papaderos et al. 1996).
 
Recent investigations (van Zee et al. 1998) reveal that BCDs 
show, unlike dIs, a compact and dense \hi-distribution peaking 
typically very close to the intensity maximum of the
starburst component (see also Taylor et al. 1994). 
Thus, if BCDs are dIs seen in active stages, the onset of a starburst 
must be accompanied by a large scale contraction of the massive 
gas-halo.
Once a burst is initiated, star formation processes as well as the 
subsequent dynamical evolution of a BCD may sensitively depend on 
the formation circumstances of a hot (few $10^6$ K) gas phase, 
a process manifesting itself in the formation of bubbles 
expanding within the ambient cold gas medium 
(Marlowe et al. 1995, Heckman et al. 1995, Bomans et al. 1997).
Although, in terms of its mass, this tenuous hot gas-phase may not be
dynamically important, the pressure it exerts on the ambient 
cold \hi-gas has obvious consequences.
The gradual increase of the volume filling factor of the hot gas phase 
within the optical BCD, i.e. the replacement of the 
\hi-gas by a warm/hot gas phase may lead to a flattening of the 
gravitational potential, thus initiating an adiabatic expansion 
of the stellar LSB-component.

Understanding the formation characteristics and evolution of this hot,
X-ray emitting gas-phase in the course of a starburst is, therefore, a basic 
requirement for gaining insights into the dynamical evolution and 
the activity status of BCDs.
% ------------------------------------------------------- FIGURE 1 ------------------
\begin{figure*}[!t]
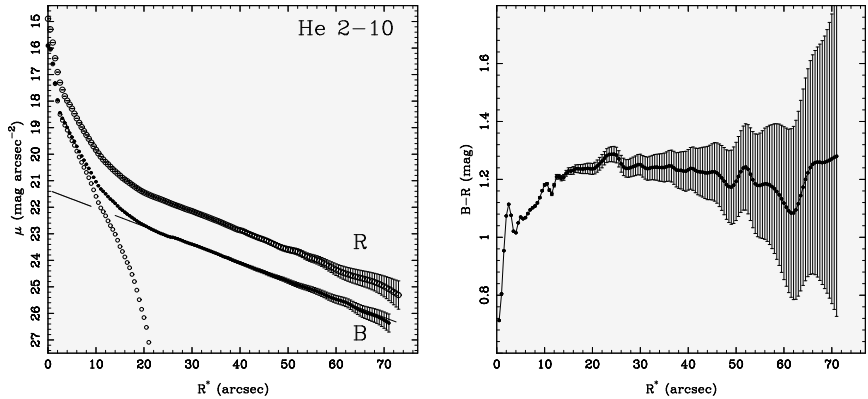

\begin{picture}(16.8,5.2)
%  To be inserted: Henize2_10f1.gif
%% \put(-0.4,0){\psfig{figure=Henize2_10f1.ps,height=5.2cm,angle=0.,clip=}}
% 
\put(6.2,0){\psfig{figure=Henize2_10f2.ps,height=5.2cm,angle=-90,clip=}}
\put(12.2,0){\psfig{figure=Henize2_10f3.ps,height=5.2cm,angle=-90.,clip=}}
\end{picture}
\caption[]{{\bf left:} B-band exposure of Henize\ 2-10 (D=8.7 Mpc) 
obtained, as also the further optical data presented below, at the 
Danish 1.54m telescope at La\ Silla. 
Contours are shown at the intensities from 18 to 26\ \sbu in steps of 0.5 mag. 
Either nucleus is separated by $\sim 9$\arcsec\ (380 pc).
The same diagram shows that for intensities $\ga 23$\ B\ \sbu\ the
light is dominated by the smooth underlying LSB component.
{\bf middle:} Surface brightness profiles of Henize\ 2-10 in B and R. 
Note the exponential intensity decrease for radii $>30$\arcsec.
Profile decomposition into the exponential component (solid line) and the 
starburst-luminosity in excess of the LSB component (small circles) is shown for the
B-band profile. The absolute B magnitudes of the underlying LSB component and of the 
starburst component within the 25\ \sbu\ isophote amount to --16.16 mag and --16.17 mag, 
respectively, i.e. either component are nearly equally luminous.
{\bf right:} Radial B--R profile of Henize\ 2-10 derived by subtraction of the 
R-band profile from the B-band profile. The colour of the underlying LSB-component,
$\sim 1.3$ mag, witnesses an older host galaxy underlying the regions of active star
formation.}
\label{bcd1}
\end{figure*}
In Sect.\,2 we shall briefly introduce a further paradigm of the 
formation of a hot gas-phase in a BCD on the basis of X-ray and 
optical data and in Sect.3 we discuss links between the structure 
of the stellar LSB-component and the activity of a BCD 
as it imprints its integrated X-ray properties.
% ------------------------------------------------------------- FIGURE 2 -------------
\begin{figure*}[!] 
\begin{picture}(16.8,6.25)
%! to be inserted: Henize2_10f4.gif
%% \put(1,0){\psfig{figure=Henize2_10f4.ps,height=6.25cm,angle=0.,clip=}}
%! to be inserted: Henize2_10f5.gif
%% \put(9,0){\psfig{figure=Henize2_10f5.ps,height=6.25cm,angle=0,clip=}}
\end{picture}
\caption[]{{\bf left:} ROSAT HRI exposure of Henize\ 2-10 overlaid to a \ha-map. 
The high surface brightness X-ray spot containing nearly half of the total flux
appears to be slightly shifted with respect to the western, most luminous 
starburst knot. 
The positions of the starburst knots, of which the western one contributes 
$\sim 3/4$ of the total \ha-emission, are indicated by crosses.
{\bf right:} \ha-equivalent width EW(\ha) map of the central region of Henize\ 2-10. 
The overlaid contours show the optical morphology in the B band and 
correspond to surface brightness levels from 17 to 20\ B\ \sbu\
in steps of 0.5 mag. The dashed line indicates the orientation of the elongated
$^{12}$CO component (Kobulnicki et al. 1995) protruding from the interface between 
either starburst knot to southeast.} 
\label{bcd2}
\end{figure*}
% ------------------------------------------------------------------------------------
% -------------------
\section{Henize 2-10}
% -------------------
The morphological properties of this Wolf-Rayet 
galaxy have prompted different interpretations of
its nature and of the origin of its starburst activity.
In one of them Henize\ 2-10 is regarded a typical 
BCD (Corwin et al. 1993), similar to those discussed in 
Loose \& Thuan (1986). 
On the other hand, the presence of two star-forming regions 
(Bergvall 1985, Johansson 1987) and the kinematics of its 
neutral and molecular gas (Kobulnicki et al. 1995)
led to the hypothesis that Henize\ 2-10 is a system of 
two merging dwarf galaxies.
Figure\ 1 shows that either starburst knot is
embedded within a nearly circular stellar envelope dominating 
the light for $\ga\!23$\ B\ \sbu. 
The average B--R index of this LSB component of 
$\sim$ 1.3 mag (Fig.\ 1,right) as well 
as its smooth morphology suggest an old, dynamically 
relaxed stellar population.
Its intensity distribution can be approximated by an 
exponential fitting law with a central surface brightness 
$\mu_{\rm B,0}=21.35\pm0.02$\ \sbu\ and a scale length 
$\alpha$=670$\pm$0.12 pc, i.e. structural properties 
which are typical for BCDs (cf. Papaderos et al. 1996).
% ------------------------------------------------------- FIGURE 3 -----------------------
\begin{figure*}
\begin{picture}(16.8,6.4)
%! to be inserted: Henize2_10f6.gif
%% \put(0,0){\psfig{figure=Henize2_10f6.ps,height=6.4cm,angle=-90,clip=}}
%! to be inserted: Henize2_10f7.gif
%% \put(9.2,0){\psfig{figure=Henize2_10f7.ps,height=6.4cm,angle=-90,clip=}}
\end{picture}
\caption[]{{\bf left:} The B-R map of Henize 2-10 shown in the range between 
0.9 and 1.07 mag. The overlayed contours delineate the morphology of the BCD 
in the B-band and correspond to intensities from 17 to 22 B \sbu\ in steps of 0.5 mag.
The complex colour distribution of the system is obvious. 
From the main starburst knot (W) proceeds a blue (B--R$\sim$0.9 mag) curved 
pattern to the west. The origin of this feature which does not coincide 
with any conspicuous enhancement in EW(\ha) (cf. right diagram), 
is currently unter investigation. 
A conceivable possibility is that it is due to an ensemble
of aging stellar clusters formed in the past.
A further, broad blue (B--R$\la$0.67 mag) region, extending 
$\sim 7$\arcsec\ northeast from the eastern starburst knot (E)  
is visible.
{\bf right:} EW(\ha)-contours superposed on the B--R map. The position of either 
nucleus is indicated by the crosses. The central starburst component (W) coincides 
with the region with the highest EW(\ha) comprising a number of local maxima on
scales of $\sim 10$\arcsec. From this active star-forming region (cf. Conti \& Vacca 1994) 
proceeds a supershell (sw) out to $\sim 560$ pc to the southwest direction. 
It is obvious that the cavity delineated by the eastern supershell (ne) 
coincides with the extended blue region in the vicinity of the starburst knot E.
The component designated C shows a B--R colour of $\sim 1.5$ mag, i.e. is by 
0.2 mag reeder than the average value of the underlying stellar 
population. This feature is located close to the northern tip of the 
CO-complex discovered by Kobulnicky et al. (1995) and may be attributed to 
enhanced intrinsic absorption.}  
\label{bcd3}
\end{figure*}
% ----------------------------------------------------------------------------
This is also the case for the concentration index CI=0.88 
which is close to the expected value for a host galaxy with 
the absolute luminosity of that of Henize\ 2-10.
On the other hand, the disturbed B--R morphology 
(Fig.\ 3,left) and the evidence described above 
suggest that He\ 2-10 may be a dynamically perturbed BCD.

A thermal bremsstrahlung fit to the PSPC-spectrum of Henize\ 2-10 
yields an average plasma temperature kT=(0.49\mima{0.18}{0.29}) keV 
and an intrinsic 0.1--2.4 keV luminosity of 
\lx=(7.6\mima{3.3}{9.3})$\times 10^{39}$ \ergsec\ 
in agreement with the value log(L$_{\rm X}$/\ergsec)=40--41
derived by Hensler et al. (1997) and Stevens \& Strickland (1998).
As HRI-maps reveal (Fig.\ 2,left) roughly one half of the 
X-ray emission is contributed by a compact high surface brightness 
source, being probably located slightly offset from the most luminous
starburst knot, close to the region with low \hi-surface density 
discovered by Kobulnicky et al. (1995; cf. their Fig.\ 10).
Given, however, the positional uncertainty of $\sim 6$\arcsec\ 
in HRI-maps a further check is required. 
The remaining X-ray emission extends on scales comparable to 
those of the \ha-emission, the faint outskirt of the latter
being visible out to $\sim$0\farcm 5 (1.3 kpc) from the nuclear 
region.

The \ha-equivalent width map of the central region of Henize\ 2-10 
(Fig.\ 2,right) suggests a kinematically perturbed ISM 
on kpc-scales with two marked supershells expanding from either starburst 
knot roughly perpendicular to the plane of the CO-complex.
The cavity delineated by the eastern supershell coincides with an 
extended blue region with an average B--R index 
$\la 0.67$ mag. This feature being centered $\sim\!3$\arcsec\
northeast from the eastern starburst knot does not seem to 
be associated with any conspicuous local enhancement of the
stellar background (cf. Sauvage et al. 1997, Beck et al. 1997).
An analysis in progress (Papaderos et al. 1998a) 
focusses on the nature of the eastern shell, 
in particular the question of whether it may be driven 
by a mixture of warm and hot X-ray emitting gas 
inflating its interior. The measured B--R index
is consistent with the values between 0.48 and 1.1 mag 
predicted by Kr\"{u}ger (1992) for a photoionized 
gaseous continuum with metallicities of 1/20$\,Z_{\odot}$ 
and $Z_{\odot}$, respectively.\\ 
These results show that Henize\ 2-10 is a well suited laboratory 
for studying the interplay between the cold, warm and hot gas phase 
of a BCD in the course of a violent starburst.
A continued investigation of this system, in particular with 
the help of spatially resolved X-ray maps, is apparently of 
great interest.
% ------------------------------------------
\section{On the effect of the LSB-component}
% ------------------------------------------
The X-ray emission as a consequence of the starburst 
phenomenon, in particular the X-ray to B-luminosity ratio 
\lxb, may hold important information on the activity status 
of starburst galaxies (cf. Fricke \& Papaderos, these proceedings).
A preliminary analysis of ROSAT-detected BCDs (Papaderos et al. 1998b)
indicates that among them the more metal-rich objects follow, similar to 
colliding starburst galaxies, a trend of increasing 
\lxb\ with increasing \lirb-ratio. 
Thus, in certain evolutionary stages the \lxb-ratio 
may be regarded a measure of the burst strength.
This, however, may not be the case in a late starburst age,
when most of the X-ray emitting gas formed in a BCD inflates 
the galactic halo where, through adiabatic expansion, 
it may obtain spectral properties and a surface brightness
comparable to those of the diffuse X-ray background. 
NGC\ 1705 (Hensler et al. 1998, cf. their Figs.\ 1\&2) 
may be considered an example of this later starburst phase.

Next we shall comment on one of the questions posed in Sect.\,1:
how the structural properties of the underlying stellar component
in a BCD may influence the starburst phenomenon.
In Fig.\,4 (top) the \lxb-ratio for a sample of 
BCDs is compared with the exponential scale 
length $\alpha$ of their LSB component. 
In the lower diagram \lxb\ is correlated with the 
ratio of the central luminosity density of the 
LSB component of BCDs with that of dIs, $l_{\rm 0,BCD}/l_{\rm 0,dI}$.
Both diagrams show a trend for increasing \lxb-ratio
with decreasing $\alpha$ and increasing $l_{\rm 0,BCD}/l_{\rm 0,dI}$. 
From the latter trends and Fig.\,8 in Papaderos et al. (1996) 
follows that the most compact and most mass-poor BCDs show the highest \lxb-ratio.
% ----------------------------------------------- FIGURE 4 -------------------
\begin{figure}
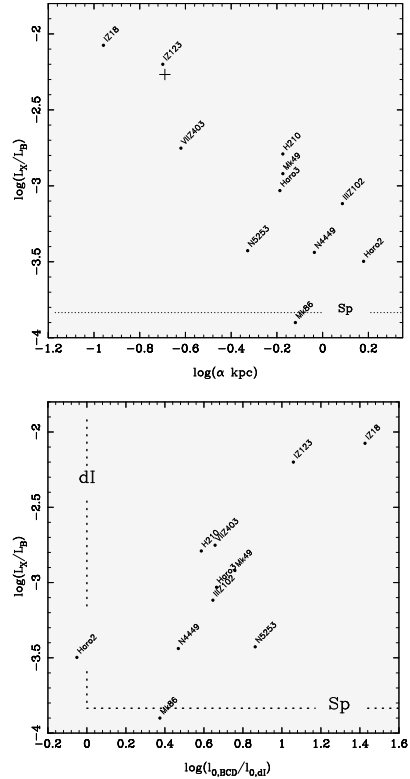

\begin{picture}(8,10.5)
\put(1.5,5.3){\psfig{figure=bcdf1.ps,height=5cm,angle=-90,clip=}}
\put(1.5,0){\psfig{figure=bcdf2.ps,height=5cm,angle=-90,clip=}}
\end{picture}
\caption[]{Comparison of the ratio of the intrinsic
0.1--2.4 keV luminosity to the integrated B-band luminosity \lxb\ 
with the structural properties of the underlying 
stellar LSB component for a sample of BCDs. 
{\bf top} \lxb\ vs. exponential scale length and {\bf bottom}
vs. ratio of the central luminosity density of the 
LSB component of BCDs to dIs (vertical dotted line).
For the latter type of dwarfs we assumed a typical value of 
23\ B\ \sbu\ and 1 kpc for their central surface brightness 
and exponential scale length, respectively.
In both diagrams the average \lxb-ratio --3.8 for isolated spiral galaxies 
(Read et al. 1997) is shown by the horizontal dotted line.}
\label{bcd4}
\end{figure}
% ------------------------------------------------------------------------------
Stevens \& Strickland (1998) remarked that a sudden
increase in the \lxb-ratio is to be expected in an 
instantaneous burst with age $>3$ Myr. 
By contrast, enduring star-formation is not expected 
to lead to such a strong increase in \lxb.
Combining all latter lines of evidence 
with the \lxb-\lirb-trend the present results may be 
interpreted as follows:
bursts igniting in compact and less-massive BCDs 
are more violent than in the more massive ones, 
in the sense that they must be shorter 
and characterized by a higher burst parameter.

\vspace*{-2mm}
\begin{acknowledgements}
This work was supported by the Deutsche Agentur f\"{u}r Raumfahrtangelegenheiten 
(DARA) grant 50\ OR\ 9407\ 6.
\end{acknowledgements}


\vspace*{-4mm}
\begin{thebibliography}{}

\vv
\bibitem{} Beck, S.C., Kelly, D.M., Lacy, J.H. 1997, AJ 114, 585

\vv
\bibitem{} Bergvall, N. 1985, A\&A 146, 269

\vv
\bibitem{} Bomans, D.J., Chu, Y.-H., Hopp, U. 1997, AJ 113, 1678

\vv
\bibitem{} Conti, P.S., Vacca, W.D. 1994, ApJ 423, L97

\vv
\bibitem{} Comte, G., et al. 1996, in {\sl In the Early Universe with the VLT}, ed. J.\ Bergeron, 127  

\vv
\bibitem{} Corbin, M.R., Korista, K.T., Vacca, W.D. 1993, AJ 105, 1313

\vv
\bibitem{} Heckman, T.M. et al., 1995, ApJ 448, 98

\vv
\bibitem{} Hensler, G., et al. 1998 ApJ 502, L17

\vv
\bibitem{} Hensler, G., Dickow, R., Junkes, N. 1997, Rev.Mex.A.A. 6, 90

\vv
\bibitem{} Johansson, L. 1987, A\&A 182, 179

\vspace*{-0.1mm}
\bibitem{} Kobulnicky, H.A., et al. 1995, AJ 110, 116

\vv
\bibitem{} Kr\"{u}ger, H. 1992, PhD thesis, G\"{o}ttingen

\vv
\bibitem{} Loose, H.-H., Fricke, K.J. 1981, in ESO Workshop: {\sl The Most Massive Stars},
eds. S.D.\ Odorico, D.\ Baade, K.\ Kj\"{a}r, 269

\vv
\bibitem{} Loose, H.-H., Thuan, T.X. 1986, in {\sl Star-Forming Dwarf Galaxies},
eds. D.\,Kunth, T.X.\,Thuan, T.T.\,Van, 73

\vv
\bibitem{} Marlowe, A.T., et al. 1995, ApJ 438, 563

\vv
\bibitem{} Marlowe, A.T., et al. 1997, ApJS 112, 285

\vv
\bibitem{} Meurer, G.R., et al. 1998, MNRAS, in press

\vv
\bibitem{} Papaderos, P. et al. 1996, A\&A 314, 59

\vv
\bibitem{} Papaderos, P., Fricke, K.J., Thuan, T.X. 1998ab, in prep.

\vv
\bibitem{} Read,\ A.M., et al. 1997, MNRAS\ 277,\ 397

\vv
\bibitem{} Sauvage, M., Thuan, T.X., Lagage, P.O. 1997, A\&A 325, 98

\vv
\bibitem{} Stevens, I.R., Strickland, D.K. 1998, MNRAS 258, 334

\vv
\bibitem{} Taylor, C.L., et al. 1994, AJ 107, 971

\vv
\bibitem{} Thuan, T.X. 1991, in {\sl Massive Stars in Starbursts}, 
eds. C.\,Leitherer et al., 183

\vv
\bibitem{} van Zee, L., Skillman, E.D., Salzer, J.J. 1998, AJ in press
\end{thebibliography}
\end{document}